\documentclass[AMA,Times1COL]{WileyNJDv5} 
\usepackage{subcaption}
\usepackage{comment}

\articletype{Under review}%

\received{Date Month Year}
\revised{Date Month Year}
\accepted{Date Month Year}
\journal{Journal of Software Evolution and Process}
\volume{00}
\copyyear{2024}
\startpage{1}

\begin{document}

\title{Which Code Statements Implement\\Privacy Behaviors in Android Applications?}

\author[1]{Chia-Yi Su}

\author[2]{Aakash Bansal}

\author[3]{Vijayanta Jain}

\author[3]{Sepideh Ghanavati}

\author[4]{Sai Teja Peddinti}

\author[1]{Collin McMillan}

\authormark{Su \textsc{et al.}}
\titlemark{Which Code Statements Implement Privacy Behaviors in Android Applications?}

\address[1]{\orgname{University of Notre Dame}, \orgaddress{\state{IN}, \country{USA}}}

\address[2]{\orgname{Louisiana State University}, \orgaddress{\state{LA}, \country{USA}}}

\address[3]{\orgname{University of Maine}, \orgaddress{\state{ME}, \country{USA}}}

\address[4]{\orgname{Google}, \orgaddress{\state{CA}, \country{USA}}}

\corres{Corresponding author Chia-Yi Su, \email{csu3@nd.edu}}

\authormark{Su \textsc{et al.}}
\titlemark{Which Code Statements Implement Privacy Behaviors in Android Applications?}

\corres{Corresponding author Chia-Yi Su, Holy Cross Dr, Notre Dame, 46556, Indiana, USA. \email{csu3@nd.edu}}

\presentaddress{Holy Cross Dr, Notre Dame, 46556, Indiana, USA.}


\abstract[Abstract]{A ``privacy behavior'' in software is an action where the software uses personal information for a service or a feature, such as a website using location to provide content relevant to a user.  Programmers are required by regulations or application stores to provide privacy notices and labels describing these privacy behaviors. Although many tools and research prototypes have been developed to help programmers generate these notices by analyzing the source code, these approaches are often fairly coarse-grained (i.e., at the level of whole methods or files, rather than at the statement level). But this is not necessarily how privacy behaviors exist in code. Privacy behaviors are embedded in specific statements in code. Current literature does not examine what statements programmers see as most important, how consistent these views are, or how to detect them. In this paper, we conduct an empirical study to examine which statements programmers view as most-related to privacy behaviors. We find that expression statements that make function calls are most associated with privacy behaviors, while the type of privacy label has little effect on the attributes of the selected statements. We then propose an approach to automatically detect these privacy-relevant statements by fine-tuning three large language models with the data from the study. We observe that the agreement between our approach and participants is comparable to or higher than an agreement between two participants. Our study and detection approach can help programmers understand which statements in code affect privacy in mobile applications.}


\jnlcitation{\cname{%
\author{Chia-Yi Su},
\author{Aakash Bansal},
\author{Vijayanta Jain},
\author{Sepideh Ghanavati},
\author{Sai Teja Peddinti},and
\author{Collin McMillan}
}.
\ctitle{Which Code Statements Implement Privacy Behaviors in Android Applications?} \cjournal{\it J Software: Evolution and Process} \cvol{2021;00(00):1--18}.}

\maketitle

\section{Introduction}
\label{sec:introduction}

Privacy notices, which are mandated by regulations~\cite{CCPA, GDPR}, are an important artifact for any software because they inform users how their personal information will be used and for what purpose. Recently, app stores have also required programmers to provide privacy ``nutrition'' labels (or just privacy labels) alongside the privacy notices/policies, since privacy policies are often riddled with legal jargon, are challenging to understand, and are not readable on small screens~\cite{AppStoreLabels, PlayLabels, korunovska2020challenges, kelley2009nutrition}. These privacy labels are pre-defined templates that succinctly describe privacy behaviors in terms of `practice' i.e., how user data is used, and `purpose' i.e., why it is used~\cite{jain2022pact}. Labels can be challenging to create, especially for small development teams, either due to difficulty in understanding the privacy behaviors or lack of communication between teams \cite{li2022understanding}.

Recently, a few research tools and prototypes have been developed to generate accurate privacy labels, by either summarizing privacy policies~\cite{jain2023atlas} or providing programmers with a questionnaire~\cite{gardner2022wiz}. While helpful, these approaches are often not linked with the source code, therefore, the labels become outdated as the code changes~\cite{li2022understanding}, resulting in ``inconsistencies'' arising between the app behavior and the privacy labels. To address this, some studies have proposed a coarse-grained analysis of source code, such as only analyzing API calls~\cite{zimmeck2021privacyflash} or analyzing multiple methods in a call graph~\cite{jain2022pact}. However, the coarse-grained analysis of source code may not provide enough information to help programmers either update privacy labels or resolve any conflicts between labels and code~\cite{jain2023}.

\begin{figure}[!t]
    \centering
    \includegraphics[width=0.65\textwidth]{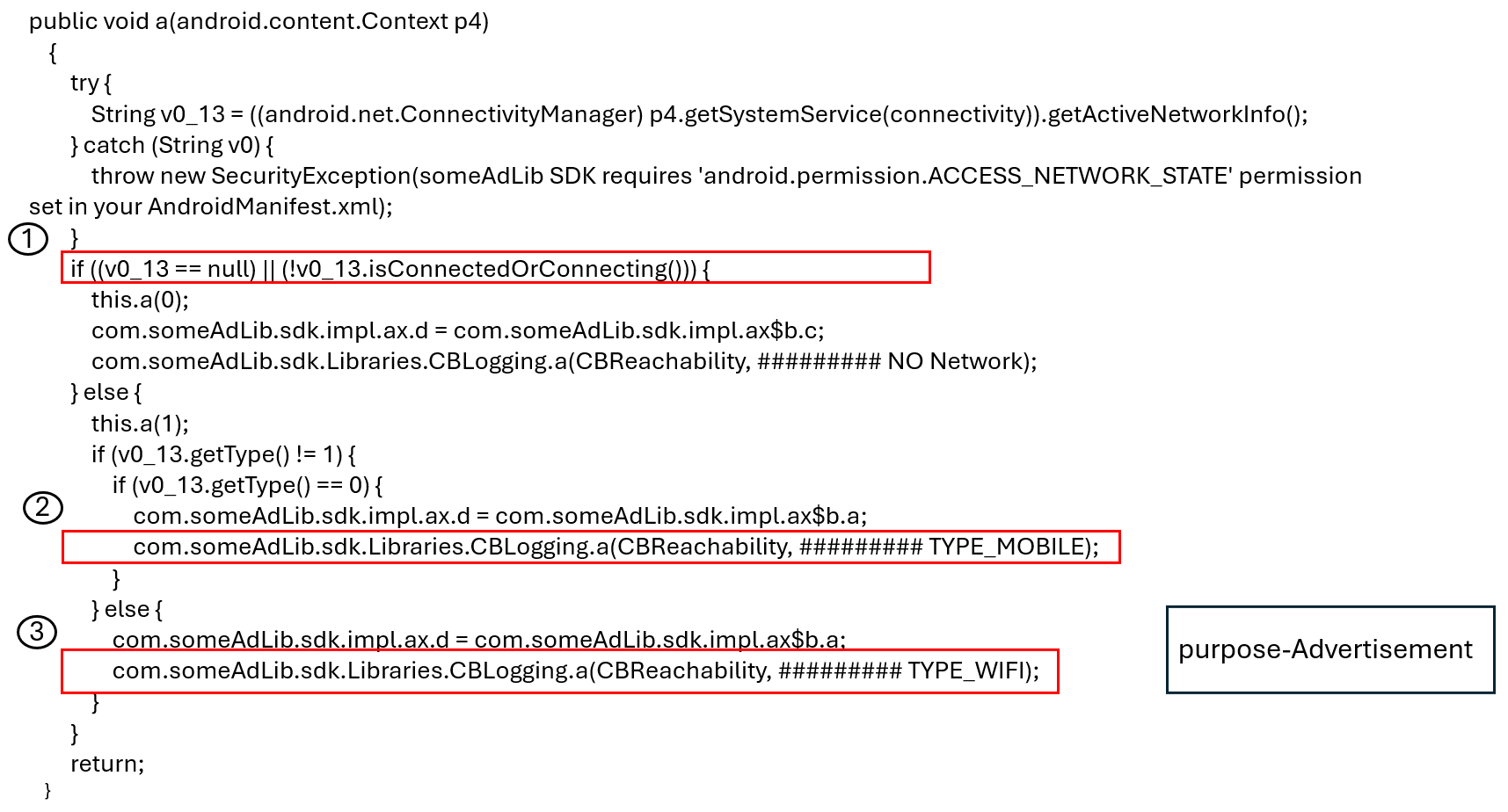}
    \vspace{-1mm}
    \caption{A motivating example of how "fine-grained" analysis highlights the embedded privacy behavior. The numbers imply the relevance of the statement to the `Advertisement' label in the `purpose' category in a decreasing order.}
    \label{fig:motivate}
    \vspace{-5mm}
\end{figure}

A barrier to better detection of privacy behaviors is that current literature does not clearly explain which code statements are likely to be involved in privacy behaviors.  Current approaches to help programmers identify and fix privacy problems would be improved if they were more fine-grained~\cite{jain2023}. ``Fine-grained'' in this context means identifying individual code statements that are involved in privacy behaviors rather than whole methods or files. A code statement is a syntactic unit in source code that executes one specific action. These actions include (but are not limited to) declarations, expressions, and iterations~\cite{nilsson2009declarative}.  These actions can be categorized as a syntactic label or a node in the Abstract Syntax Tree (AST) of the source code~\cite{neamtiu2005understanding,collard2013srcml}. But current literature is unclear about which of these statements are likely to affect privacy.

Fine-grained analysis of the source code helps better understand the privacy label, because it shows the statements related to the specific label. Figure~\ref{fig:motivate} shows a motivating example where the three highlighted statements are related to the privacy label `Advertisement' in the `purpose' category. These statements are relevant for the `Advertisement' label because the network information is saved by a third-party advertisement SDK. When the source code is modified, the programmer can reassess the privacy label based on the highlighted statements. On the other hand, coarse-grained analysis would require programmers to comprehend the entire source code.

In this paper, we conduct an empirical study to investigate which code statements are associated with privacy behaviors. We study the syntactic nature of these statements in the context of the method. We start with a public dataset ADPAc~\cite{jain2022pact} of $\sim$5,200 methods from Android apps involved in privacy behaviors and $\sim$14,000 privacy labels. Each label helps identify either the `practice' or `purpose' associated with the privacy behavior of these methods. We design and conduct a human study with 18 programmers that identified specific code statements in 2,426 of those methods that are relevant to a given privacy label. We only focus on `purpose' privacy labels in the current iteration, because we want to maximize the representation per label with a limited number of study participants. We observe that participants found statements that make function calls, specifically expression statements that augment or save the data from function calls to local variables to be most relevant to privacy behaviors. We found that the type of privacy label has very little effect on the distribution of relevant statements.

We then present an approach for automatically identifying these statements given the code and the privacy labels using Large Language Models (LLMs). We evaluate our approach using three LLMs -- including two commercial models, and one smaller, domain-specific academic model. Since each study participant identified the three most relevant statements in each method for a label, we compute agreement between our automated approach and the study participants, both when neglecting and considering the order of  statements' relevance. We found that there is a high degree of subjectivity in human judgment about which statements are important for a privacy label but the overall (non-order specific) agreement between the model predictions and participants is comparable to those between two participants. Surprisingly, we found that the small domain-specific language model outperformed the commercial language models for this task in several situations. When considering the order of the relevant statements, the agreement between two participants significantly drops, and surprisingly there is higher agreement between the model predictions and participants than between two participants.  Overall, the high degree of subjectivity in human judgement shows the difficulty of the task, and high agreement between model predictions and the participants shows that our automatic methods can be used to identify statements relevant to privacy labels without the extra cost (both time and money) of hiring privacy experts.

This paper makes the following key contributions:
\vspace{-0.35cm}
\begin{enumerate}
    \item We perform an empirical study with 18 Java programmers to map privacy labels to code statements and release a new dataset of 2,426 Android app methods with label-to-statement annotations (Section~\ref{sec:study}).
    \item We study the composition of the statements labeled in our empirical study to determine the attributes of code statements that are related to privacy behaviors and to specific `purpose' privacy labels (Section~\ref{sec:Analysis}).
    \item We design an LLM-based approach to automatically detect statements in source code that contain privacy behaviors (Section~\ref{sec:approach}).
    \item We evaluate our approach using three LLMs, and show that the agreement between the model predictions and participants is comparable or better than between two participants (Section~\ref{sec:eval}).
\end{enumerate}

\section{Related Work}
In this section, we discuss key related work, such as issues with privacy notices, approaches that identify inconsistencies between app privacy behaviors and notices, and tools to generate privacy notices - including privacy labels; Figure~\ref{tab:background} summarizes this research landscape. We also discuss the use of large language models for code comprehension.

\begin{figure}[!b]
    \vspace{-0.5cm}
		\centering
        \small
		\begin{tabular}{p{4.3cm}|p{0.26cm}|p{0.26cm}|p{0.26cm}|p{0.26cm}|p{0.26cm}|p{0.26cm}|}
			& T     & N          &I          &D             & A          & C                \\
            \hline
            \textcolor{white}{*}ScanDal (2012)~\cite{kim2012scandal}					&   &  	&x    &    &x   & \\
            \textcolor{white}{*}WHYPER (2013)~\cite{pandita2013whyper}                  &   &  &x    &x    &   & \\
            \textcolor{white}{*}AppIntent (2013)~\cite{yang2013appintent}                  &   &  &x    &    &x   & \\
            \textcolor{white}{*}CHABADA(2013)~\cite{gorla2014checking}                  &   &  	&x    &x    &x   & \\
            \textcolor{white}{*}PAGE (2014)~\cite{rowan2014encouraging}				    &x   &   &    &x    &   & \\
            \textcolor{white}{*}PrivacyInformer (2014)~\cite{miao2014privacyinformer}   &x   &   &    &x    &   &x \\
            \textcolor{white}{*}DroidJust (2015)~\cite{chen2015droidjust}				&x   &   &x    &    &x   & \\
			\textcolor{white}{*}AutoPPG (2016)~\cite{yu2016toward}						&x   &   &    &    &x   &x \\
            \textcolor{white}{*}PVDetector (2016)~\cite{slavin2016pvdetector}            &x   &   &x    &    &x   & \\
            \textcolor{white}{*}Polidroid-AS (2017)~\cite{Slavin2017PoliDroidASA}	            &x   &  	&    &x    &   &x \\
            \textcolor{white}{*}GUILeak (2017)~\cite{wang2018guileak}				    &   &   &x    &x    &x   & \\
            \textcolor{white}{*}HybriDroid (2017)~\cite{chen2017automatic}				&x   &  	&x    &x    &   &x \\
            \textcolor{white}{*}BridgeTaint (2018)~\cite{bai2018bridgetaint}	        &   &  	&x    &    &   &x \\
            \textcolor{white}{*}CLAP (2018)~\cite{liu2018mining}						&   &x  	&    &x    &   & \\
            \textcolor{white}{*}PPChecker (2018)~\cite{yu2018ppchecker}	                &   & 	&x    &    &x   &x \\
            \textcolor{white}{*}LeakDoctor (2019)~\cite {wang2019leakdoctor}            &   &  	&x    &    &   &x \\
            \textcolor{white}{*}Android-Privacy (2020)~\cite{feichtner2020understanding}&   &x   &x   &    &   &x \\
            \textcolor{white}{*}PriVot (20201)~\cite{atapattu2021sensitive}             &   &x   &    &    &x   & \\
            \textcolor{white}{*}PrivacyFlash Pro (2021)~\cite{zimmeck2021privacyflash}	&x   &  	&    &    &x   &x\\
            \textcolor{white}{*}PriGen (2021)~\cite{jain2021prigen}                     &   &x  	&    &    &   &x \\
            \textcolor{white}{*}PAcT (2022)-\cite{jain2022pact}                         &   &x  	&    &    &   &x \\
            \textcolor{white}{*}Jain et al. (2023)~\cite{jain2023}                           &   &x  	&    &    &   &x \\
            (this paper)                                                                &  	&x    &    &   &    &x \\
		\end{tabular}
	\caption{Snapshot of recent research approaches for generating and analyzing privacy notices. $T$ denotes the use of templates.  $N$ denotes the use of Neural Networks. Column $I$ denotes approaches for inconsistency detection. $D$ denotes approaches that analyze the developer's descriptions and answers to questionnaires. Column $A$ denotes approaches that analyze API calls.  $C$ denotes approaches that use code comprehension.}
	\label{tab:background}
\end{figure}

\subsection{Notifying Privacy Behaviors}

Early work in notifying privacy behaviors of mobile apps focused on the practice of ``Notice and Consent''~\cite{liu2014user}. In this practice, most apps relied on using long privacy policies as privacy ``notices'', and assumed users' ``consent'' if they downloaded and used the app. However, several works criticized this model~\cite{liu2014user,okoyomon2019ridiculousness,almuhimedi2015your,lin2012expectation} because they found that users are not immediately aware of these privacy trade-offs, and privacy policy did not serve as an effective medium for notice. 


One proposed solution is to notify and ask for consent every time the app uses sensitive data, as opposed to one-time consent. While effective, this approach leads to notification fatigue~\cite{utz2019informed,schaub2015design,balebako2015impact,karegar2020dilemma}. The current state of practice is an ``opt-in'' approach, where users have fine-grained control over the data they want to share with the app. Additionally, users can now change these permissions in real-time after the app is downloaded.

\subsection{Privacy Labels}
\label{sec:labels}

Privacy labels are akin to ``nutrition labels'' that provide important information about privacy and data handling in a concise manner. These were first proposed as an alternative to long privacy policies in 2009 by Kelley~\emph{et al.}~\cite{kelley2009nutrition}. More recently, academic works have increasingly argued in favor of short privacy labels~\cite{susser2019notice,reidenberg2018trustworthy}. The main arguments in favor of privacy labels are widespread non-compliance with policies, pitfalls of notice and consent, notification fatigue, and coarse-grained control of user data permissions~\cite{gardner2022wiz}.

These proposed changes were adopted by Apple in 2020 and Google in 2022 as the two major stakeholders of the mobile application marketplace~\cite{rodriguez2023comparing}. The major challenge now is maintaining the accuracy of privacy labels at all times. Developers may not always be aware of all data handling practices in their app due to third-party features such as advertisement networks. 
One solution is automatic privacy label generators that would allow app developers to keep labels updated through code changes, and version updates~\cite{jain2022pact, jain2023}.

\subsection{Automatic Generation of Privacy Labels}

Foundational work in automatic privacy label generation was based on automatic privacy policy generators such as PAGE~\cite{rowan2014encouraging} and  PrivacyInformer~\cite{miao2014privacyinformer}. These approaches used the descriptions and questionnaire answers provided by the developer to generate policies, often using a template as denoted in Figure~\ref{tab:background}. However, developers' responses and descriptions are not entirely reliable for the same reasons as discussed in Section~\ref{sec:introduction}.

In 2016, Yu~\emph{et al.}~\cite{yu2016toward} released AutoPPG, a tool for automatic generation of privacy notices using static analysis of API calls and source code. Around the same time, Slavin~\emph{et al.}~\cite{Slavin2017PoliDroidASA} released Polidroid-AS, a tool for automatic generation of privacy descriptions using both static code analysis and developer questionnaires. The short descriptions produced by Polidroid-AS are akin to privacy labels but not as comprehensive. These approaches use pre-written templates.

In 2022, Jain~\emph{et al.}~\cite{jain2022pact} proposed a neural network-based approach for the automatic generation of privacy labels by using language models to translate code snippets into correlating privacy labels. Later, they extended this work by increasing the size of code snippets to multiple methods in a call graph and finely-localizing individual statements in those methods that correspond to the predicted privacy labels~\cite{jain2023}. The essence of these works is that generating privacy labels using source code can be seen as an automatic code comprehension task.

\subsection{Inconsistency Detection}
\label{sub:Inconsistency}

Apart from automatically generating notices, several studies have also focused on detecting inconsistencies between an app's privacy behavior and its privacy notices. While most studies have focused on detecting inconsistencies in privacy policies ~\cite{yu2018ppchecker, okoyomon2019ridiculousness, zimmeck2019maps}, there have also been works on detecting inconsistencies in app descriptions~\cite{gorla2014checking}, permission rationales~\cite{liu2018large}, and even privacy labels~\cite{xiao2022lalaine}. While it is concerning to have inconsistencies for all app categories, it is alarming for apps that deal with sensitive personal data such as health~\cite{papageorgiou2018security}, financial~\cite{wang2018guileak}, or children's data, which are protected by several acts~\cite{mcdonald2013nano,reyes2017our}. 

A reason for  these inconsistencies is the re-use of privacy-sensitive code by developers from other apps. Such a scenario may cause a compounding effect for inconsistencies between privacy notices and app behavior because the developer may have re-used privacy-sensitive code without implementing the protections or disclosing the risks from the source app~\cite{watanabe2015understanding}. Tools such as HybriDroid~\cite{chen2017automatic} and LeakDoctor~\cite{wang2019leakdoctor} build models of the app and find specific regions of code that might be misaligned with the indented functionality of the app.  While these techniques exposed the underlying problem with mobile application developers' handling of user data, they also outlined the ineffectiveness of long privacy policies.

\subsection{Language Models for Code Comprehension}

Recently, there has been an emergence of language models for code comprehension. These approaches learn to model source code as a language using copious number of samples. Early work in language models for code comprehension used encoder-decoder based techniques for tasks such as code summarization~\cite{leclair2019neural,bansal2021project,bansal2023callcon}, code clone detection~\cite{ghofrani2017conceptual,hua2020fcca,meng2020deep}, code generation~\cite{xie2021improving,sun2019grammar}, and program repair~\cite{gupta2020synthesize,lutellier2020coconut}. 

The latest approach in automatic code comprehension are Large Language Models (LLMs). These are scaled-up versions of foundational language models for large-scale industrial applications. Popular industrial-scale models such as OpenAI GPT and Google Gemini are trained on a variety of data scrubbed from the web, including natural language text, source code, and user interactions. However, these approaches are closed source, i.e., their training data and fine-tuning practices are unknown. Alternatively, open-source approaches trained specifically for code such as CodeT5~\cite{wang2021codet5}, StarCoder~\cite{li2023starcoder}, CodeLLama~\cite{roziere2023code} have shown comparable performance for other code comprehension tasks such as code summarization~\cite{chia2023language}, code generation~\cite{le2022coderl}, and program repair~\cite{olausson2023self}.  

Although these LLMs may demonstrate the advanced ability on code comprehension tasks, the goal of these automatic approaches is not to compete with the human experts. Instead, these approaches are to increase the productivity of the programmers~\cite{ciniselli2024today}. In this paper, our goal is to design LLMs that predict code statements relevant to the given privacy labels.



\section{Survey Design}
\label{sec:study}

\begin{figure}[h]
    \centering
    \includegraphics[width=0.75\textwidth]{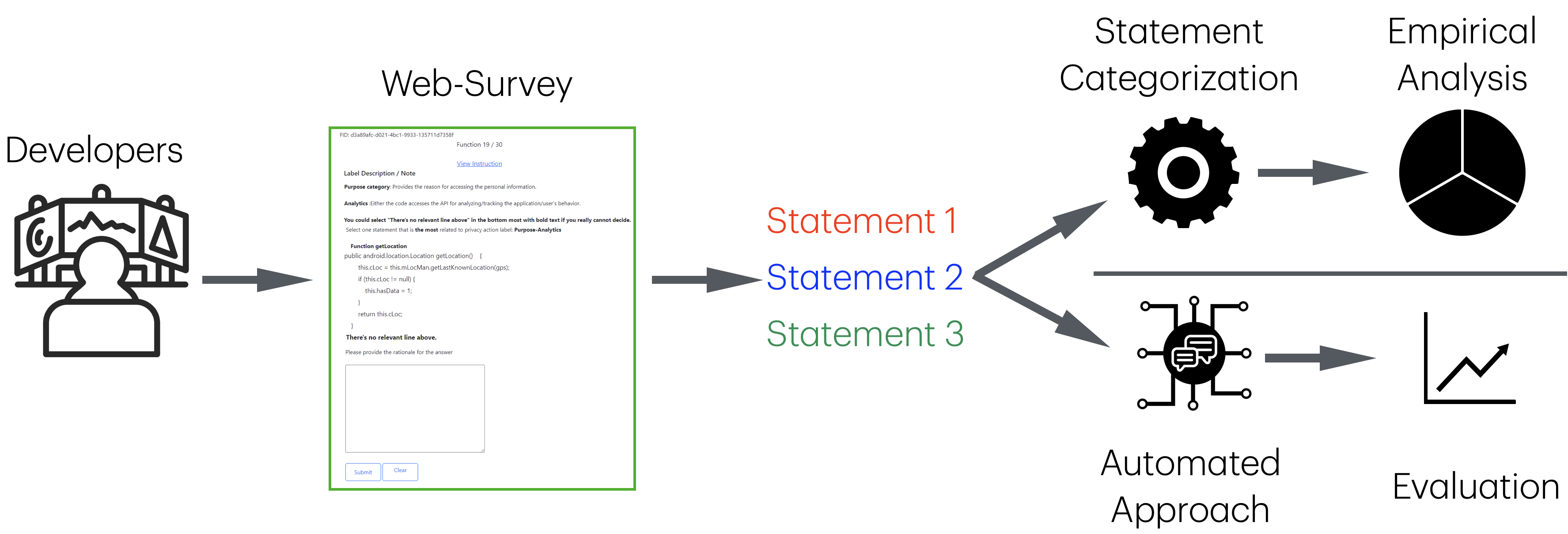}
    \vspace{-1mm}
    \caption{Overview of our study. We discuss our web-survey study and statement categorization in Section~\ref{sec:study}. We discuss the results of our empirical analysis in Section~\ref{sec:Analysis}. We discuss the features of our automated approach in Section ~\ref{sec:approach}, and the evaluation of that approach in Section~\ref{sec:eval}. }
    \label{fig:overview}
    \vspace{-5mm}
\end{figure}

Figure~\ref{fig:overview} provides an overview of the entire paper. In this section we describe the design of our web-survey experiment and statement categorization. The purpose of this experiment is to identify the attributes of code statements that programmers perceive as most relevant to privacy behaviors in software. We describe our research questions, dataset, participants, and research method in the following subsections.

\subsection{Research Questions}
\label{sub:rqs1}
We ask two Research Questions (RQs) pertaining to the ``attributes'' of code statements. We define the ``attributes'' as Abstract Syntax Tree (AST) node types, as recommended by Zhu~\emph{et al.}~\cite{zhu2015analysis} (see Section~\ref{sub:statements}). The RQs are:
\begin{description}
    \item [RQ1:] Which statement attributes are relevant to privacy behaviors, in general?
    \item [RQ2:] Which statements attributes are related to specific types of privacy labels?
\end{description}

The rationale behind RQ1 is to understand details about the code statements that affect privacy overall. These code statements are individual statements that implement privacy behavior, as defined by  the set of `purpose' privacy labels introduced by Jain~\emph{et al.}~\cite{jain2022pact} (see Section~\ref{sec:labels}). This analysis provides a fine-grained understanding of  how statement attributes affect privacy.

We further breakdown the analysis of statement attributes by specific `purpose' privacy label types in RQ2.  
This breakdown provides understanding of which statement attributes may affect specific types of privacy behaviors.


\subsection{Dataset and Privacy Labels}
\label{sub:privacylabels}
The 2,426 samples of code and privacy labels in our dataset were randomly selected from the ADPAc dataset created by Jain~\emph{et al.}~\cite{jain2022creating}.  That dataset contains around 5,200 Android methods and around 14,000 labels for those methods. The labels are also divided into two categories based on 1) `practice' -- how the information is used, and 2) `purpose' -- why is it used. We only use the privacy labels that describe the `purpose' of the privacy behavior for two reasons. The first is to increase the representation per individual privacy label in the data we collect. Since we can only recruit a limited number of participants, including more labels would decrease the representations per label. The second reason is that we use this dataset to fine-tune a language model in Section~\ref{sec:approach}. When there is more than one label per sample, training a model to predict statements given labels increases the number of variables and is outside the scope of this study. Jain~\emph{et al.} proposed four privacy labels for the `purpose' behavior~\cite{jain2022pact}:

\begin{enumerate}
\item \textbf{Advertisement:}  when the personal data is being used for advertisement services.
\item \textbf{Functionality:} when the personal data is being used for the functionality of the app.
\item \textbf{Analytics:} when the personal data is being used for analytics in or outside the app.
\item \textbf{Other:} when the personal data is being used for other/unknown purposes.
\end{enumerate}

\newpage

When a method had more than one `purpose' label, we randomly selected one of the labels. Our rationale was to keep the task simple for the participants. Since a method can be linked to several privacy behaviors, including all `purpose' labels may complicate the task by requiring the participants to find the relevant code statements to all behaviors and also decide which privacy behavior is most important. 
By providing one specific label we give the participant a well-defined task to complete in a reasonable time-frame. 

In this study, we do not use sampling methods (e.g., stratified sampling) popular in traditional machine learning, because we want to analyze the human expert annotations and use all of the limited annotated dataset to evaluate the effectiveness of training a model to capture human expert knowledge. The sampling method may complicate the results and is outside the scope of this paper.

\begin{figure}[!t]
    \centering
    \includegraphics[scale=0.45]{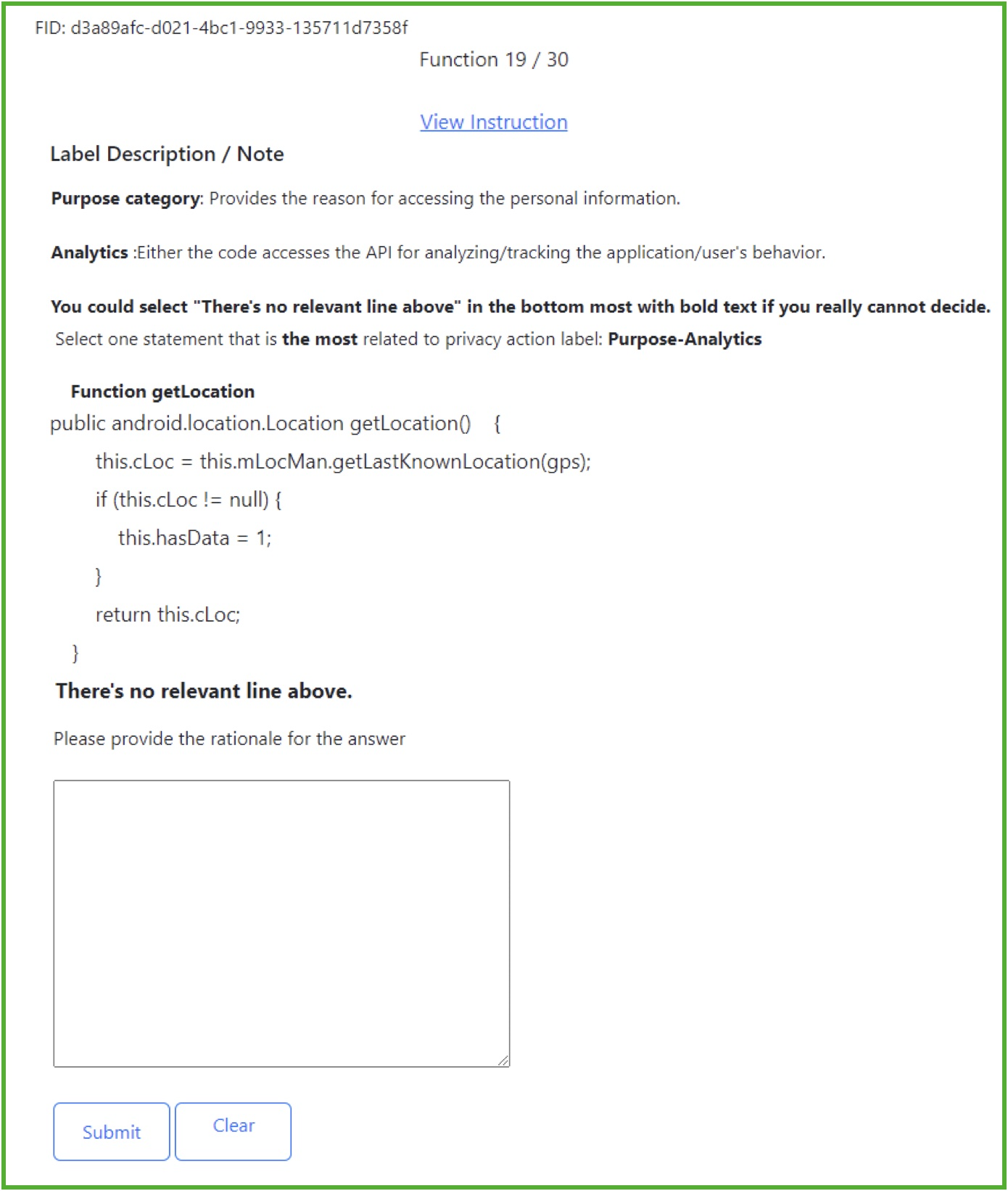}
    \vspace{-3mm}
    \caption{A snapshot from our web-survey.}
    \label{fig:survey_question}
    \vspace{-6mm}
\end{figure}

\subsection{Participants}

We used Prolific~\cite{palan2018prolific} to recruit participants for our survey. Participants were Java developers of at least 25 years of age, and held a degree in Computer Science or Engineering. We manually inspected the first session to filter participants who either provided nonsensical rationales or statement selections without content (e.g., empty lines, a line with only a curly brace). Though this manual filtering is biased by our judgment, it is necessary because, as Ghorbani~\emph{et al.}~\cite{ghorbani2023autonomy} point out, participants can now easily circumvent the standard programming screening questions~\cite{danilova2021you} with AI models such as ChatGPT and Github Copilot. We recruited 23 participants, and filtered out 5 to analyze data from 18 qualified participants.

\subsection{Research Method}
\label{sec:studymethod}
    Our key research method is a web survey in which we show programmers a source code and a privacy label for that code, then we ask them to select the top three most relevant statements in that code for that privacy label. We modeled our web survey on published experiments that collect human judgments of importance~\cite{sen2020human}. The survey begins with informed consent documents, instructions, and a sample question. An example is shown in Figure~\ref{fig:survey_question}. Note, the participant sees a privacy label, a description of that label, and source code of the corresponding method. Our task has four main steps:
    
\begin{enumerate}
\item The survey asks the participant to select one statement that the participant considers \emph{most-related} to provided privacy label. There is a text-box where the participant enters a rationale about their selection. The participant may select ``There's no relevant line above'' option if they believe that no statement is relevant.

\item In case the participant selects the no relevant line statement, the page shows a notification confirming the participant's selection. If they cancel, they can make the selection again. In case the participant selects a statement and presses submit button, the survey continues.

\item When a statement is selected, the survey shows the same code again to ask for the \emph{second-most} and \emph{third-most} relevant statements. The page highlights the \emph{first-most} relevant statement in red and the \emph{second-most} related statement in blue after selection to prevent them from selecting a statement more than once.

\item After the participant confirms that there are no more relevant statements or selects three statements, the survey proceeds to the next method.

\end{enumerate}

The end result is that for each participant, we collect a set of methods with their top three relevant statements (except in cases where the participant has indicated that no more relevant statements are present). We gave each participant a maximum of 90 minutes per session to annotate methods and some participants completed multiple sessions. Each participant was provided a randomized set of samples and due to their different paces, participants annotated a varying numbers of methods in each session.  Roughly ten percent of the methods were annotated by more than one participant. This small subset with two participants' annotations captures varying personal opinions, and may help us better evaluate the performance of our automated approach (see Section~\ref{sec:approach}) in case of disagreements. In total, 18 participants annotated 2,426 methods, of which 2,167 methods were annotated by only one participant and 259 methods were annotated by two participants.

\subsection{Statement Categorization}
\label{sub:statements}
We categorize the code statements represented by their AST nodes, as defined by Zhu~\emph{et al.}~\cite{zhu2015analysis}. There are over fifteen categories of statements extracted using the AST. See the srcML library~\cite{collard2013srcml} for exhaustive list and definitions, which we used to generate the AST. We describe seven of those categories most frequently found in our dataset:
\begin{description}
    \item [func\_call] is the statement that contains function calls, and denotes an exchange with another method.
    \item [expr\_ stmt] is the statement that contains data operators such as mathematical operations.
    \item [decl\_stmt] is the statement that declares a new variable inside the body of the method.
    \item [function\_sig] is the statement that defines the target method including name, parameters and return type.
    \item [if\_stmt] is the statement that defines conditions under which a block code of code must be executed.
    \item [else] is the statement that precedes the alternate block of code to be executed when the if conditions are not met.
    \item [return] is the statement used to exit out of the method, optionally with values.
\end{description}

Note, some of these statements supersede other statements.  For example, func\_call supersedes decl\_stmt and expr\_ stmt.  See the next section for details.

\section{Statement Analysis}
\label{sec:Analysis}
In this section, we report and discuss results of our statement analysis to answer RQ1 and RQ2 defined in the last section.

\subsection{Results for RQ1}
\begin{figure}[t]
    \centering
    \begin{tabular}{cc}
        \begin{subfigure}[b]{0.35\textwidth}
            \centering
            \includegraphics[width=\textwidth]{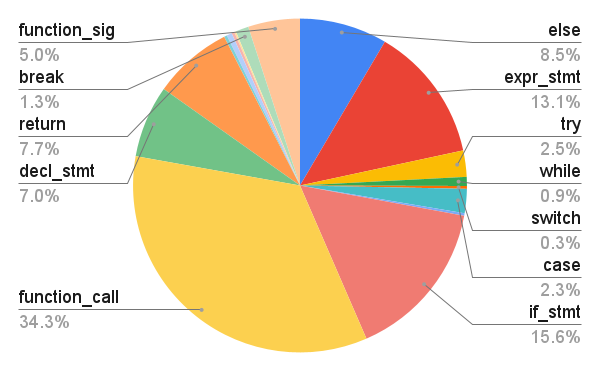}
            \caption{All Statements}
        \end{subfigure}
        \hspace{0.5cm}
        &
        \begin{subfigure}[b]{0.35\textwidth}
            \centering
            \includegraphics[width=\textwidth]{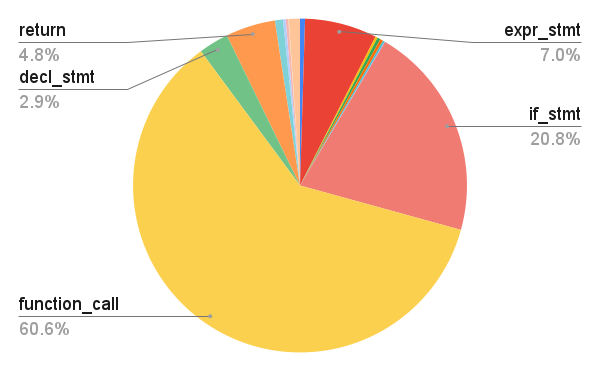}
            \caption{All Ratings}
        \end{subfigure}
        \\
        \begin{subfigure}[b]{0.35\textwidth}
            \centering
            \includegraphics[width=\textwidth]{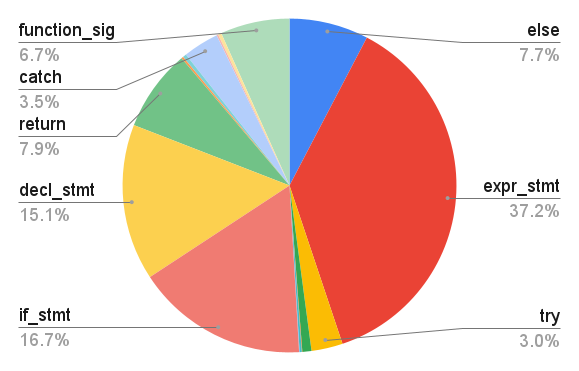}
            \caption{All Statements without func\_call}
        \end{subfigure}
        \hspace{0.5cm}
        &
        \begin{subfigure}[b]{0.35\textwidth}
            \centering
            \includegraphics[width=\textwidth]{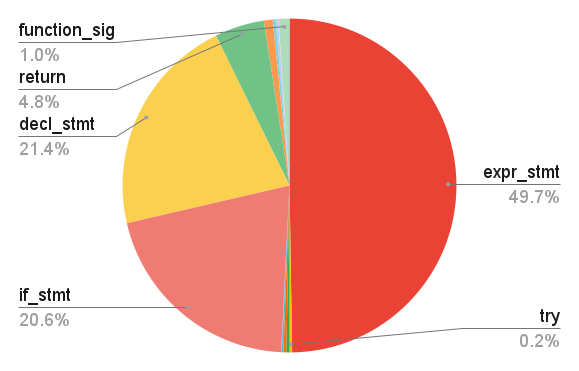}
            \caption{All Ratings without func\_call}
        \end{subfigure}
    \end{tabular}
    \vspace{-0.2cm}
    \caption{Charts showing the normalized distribution of statement categories for: (a) all statements in the methods,(b) all ratings from participants, (c) all statements without the func\_call, and (d) all ratings without the func\_call.}
    \label{fig:rq1}
    \vspace{-5mm}
\end{figure}

We summarize the results of our statement analysis for RQ1 in Figure~\ref{fig:rq1} and Table~\ref{tab:order}. In the former, we discuss overall attributes of statements and human ratings. In the latter, we report attributes based on the order in which participants selected the statements. We make three observations. 

\begin{table}[!b]
\vspace{-5mm}
\caption{Trends for order of statements selected.}
\vspace{-1mm}
\centering
\begin{tabular}{l@{\hspace{1pt}}c@{\hspace{6pt}}llll}
     & &  & \multicolumn{3}{c}{Statement Order} \\
 \multicolumn{3}{c}{}      & First & Second  & Third \\ \cline{4-6} \rule{0pt}{7pt} 
 
\multirow{4}{*}{\rotatebox{90}{Statement}}    & \multirow{4}{*}{\rotatebox{90}{Categories}} 
                                                & \multicolumn{1}{l|}{expr\_stmt}  & 49.1\%  & 49.1\% &  51.5\%     \\
                                                & & \multicolumn{1}{l|}{decl\_stmt}   & 27.7\%  & 19.5\% &  15\%   \\
                                                & & \multicolumn{1}{l|}{if\_stmt}   & 17.3\%  & 25.0\% &  19.8\%  \\
                                                & & \multicolumn{1}{l|}{return}  & 3.8\%  & 2.6\% & 8.7\%  
\end{tabular}
\label{tab:order}
\end{table}

First, we found that participants consider function calls to be most frequently associated with privacy behaviors. Comparing Figure~\ref{fig:rq1} (a) and (b), we observe that although function calls are only made in about a third of all statements, they are present in roughly 60\% of statements that are related to the privacy label. Surprisingly, \textit{if conditions} are second most likely (at 20.8\%) to be associated with a privacy behavior, over statements that augment data such as expr\_stmt or a decl\_stmt. This observation may be due to the way Zhu~\emph{et al.}~\cite{zhu2015analysis} categorize statements -- specifically that if a function call is made within an expr\_stmt or a decl\_stmt, these are categorized as func\_call statements. As function calls overwhelmingly dominate the categorization and supersede other categorizations, we further analyze this distribution by disabling the categorization of func\_call statements.

When not considering function calls as a separate statement category, expr\_stmt is the category most frequently picked by the participant as being relevant to privacy labels. In Figure~\ref{fig:rq1} (c), we present the new distribution of statements after disabling the func\_call extraction. In Figure~\ref{fig:rq1} (d), we present the corresponding distribution of statements selected by the participants. The change in distribution suggests that a majority of the func\_call statements are expr\_stmt or decl\_stmt. These results show that expr\_stmt at 49.7\% is the most popular category relevant to privacy labels. This is expected as an expr\_stmt may augment the data exchanged through a function call using mathematical operators or save the data to local variables. Next is decl\_stmt at 21.4\%, which is also expected as a decl\_stmt can be used to copy the data from a function call or send data outside the method with a function call. Surprisingly, if\_stmt closely follows decl\_stmt at 20.6\%, which is a conditional statement. 


We also asked participants to pick statements in the order of relevancy to privacy labels. In Table~\ref{tab:order}, we report the distribution of the top four statements from Figure~\ref{fig:rq1} (d), for the order in which they were selected by the participant. exp\_stmt is the most popular category irrespective of the order of relevance. We found that a participant is more likely to pick a decl\_stmt first compared to an if\_stmt, which is more likely to be picked second or third in the order of relevancy to the privacy label. We noticed that considering the first most relevant statement, decl\_stmt is picked 27.7\% of the time, compared to 17.3\% for if\_stmt. In contrast, when considering the second and third most relevant statements, an if\_stmt is picked more frequently (25\% and 19.8\%) than a decl\_stmt (19.5\% and 15\%). We posit that participants pick if\_stmt because the conditions are responsible for the execution of other statements like expr\_stmt and decl\_stmt, which are more directly related to privacy behaviors. Therefore, participants may be less likely to pick if\_stmt as~\emph{first-most} relevant.

Overall, we found that participants associate privacy labels first and foremost with statements that make function calls and specifically those that modify, save, or share data using function calls. So far we discussed the overall distribution of statements based on the way they are categorized and the order in which they were picked. However, there is another factor -- the four privacy labels, the effect of which we analyze next.

\subsection{Results for RQ2}
\begin{figure*}[!t]
    \centering
    \begin{tabular}{cc}
        \begin{subfigure}[b]{0.35\textwidth}
            \centering
            \includegraphics[width=\textwidth]{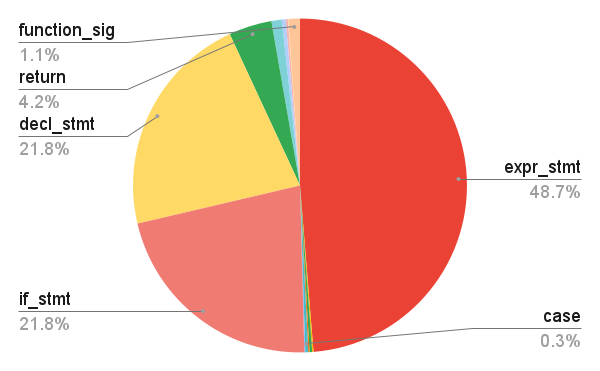}
            \caption{Advertisement}
        \end{subfigure}
        \hspace{0.5cm}
        &
        \begin{subfigure}[b]{0.35\textwidth}
            \centering
            \includegraphics[width=\textwidth]{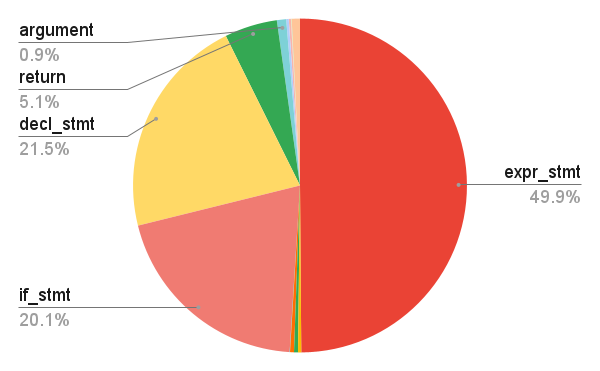}
            \caption{Functionality}
        \end{subfigure}
        \\
        \begin{subfigure}[b]{0.35\textwidth}
            \centering
            \includegraphics[width=\textwidth]{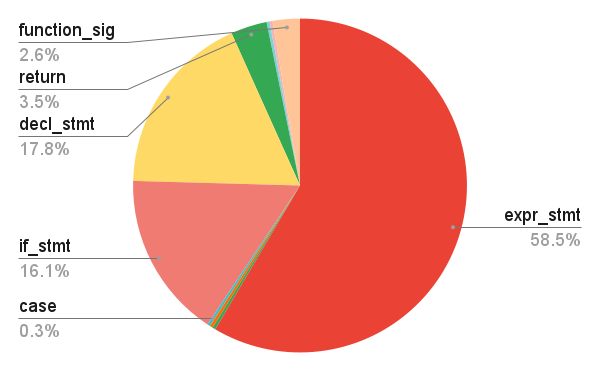}
            \caption{Analytics}
        \end{subfigure}
        \hspace{0.5cm}
        &
        \begin{subfigure}[b]{0.35\textwidth}
            \centering
            \includegraphics[width=\textwidth]{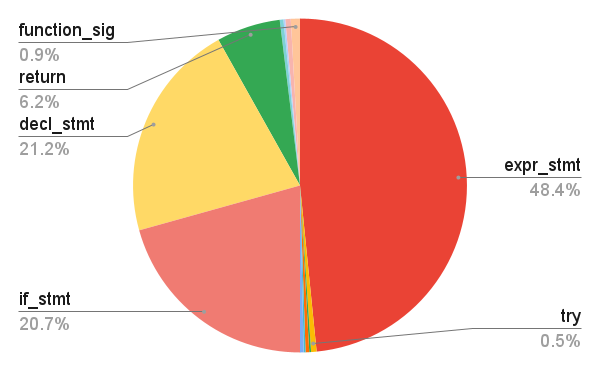}
            \caption{Other}
        \end{subfigure}
    \end{tabular}
    \vspace{-2mm}
    \caption{Charts showing the normalized distribution of all participant ratings across statement categories (without the func\_call) for the four ``purpose'' privacy behaviors.}
    \label{fig:rq2}
    \vspace{-5mm}
\end{figure*}

We summarize the results of statement analysis for the four `purpose' privacy labels in Figure~\ref{fig:rq2}. We found that the type of privacy label has very little effect on the distribution of statements. In Figure~\ref{fig:rq2}(c), we observe that participants chose expr\_stmt for an even higher number of samples when the privacy label was `Analytics', compared to the other three privacy labels. However, the order of the top most important statement categories remains unaffected by privacy labels. A potential reason for the lack of privacy label impact may be because in some instances, part of the `purpose' label information may exist outside the method that the participant does not have visibility into. For example, for a method with the `Advertisement' purpose label, the participant may see that user data is being sent outside the method via function calls in multiple expression or declaration statements and they may annotate one of these statements as being related to advertisement. However, the participant does not have access to the other methods that consume this sent information, so they may not be able to accurately compare unannotated statements for relevance to the given privacy label. Overall, these observations are consistent with the findings from RQ1, that the expr\_stmt is most likely to be associated with privacy behaviors, followed by decl\_stmt, if\_stmt, and return statements in that order, irrespective of the privacy label.

\subsection{Implications of Statements Analysis}
\label{sec:implication}

Overall, we summarize two key implications from our analysis. First, we observe that expr\_stmt is most frequently associated with the privacy behaviors, followed by decl\_stmt and if\_stmt. This observation may prompt stakeholders to examine these statements while creating privacy labels. Specifically, they may focus on the statements that involve function calls or mathematical operations, because those operations are more likely to implement privacy behaviors such as data exchange. As for if\_stmt, we recommend the stakeholders to examine the statements inside the `if' block because these statements are more likely to be involved in privacy behaviors. 

Another observation from this analysis is that the specific privacy category does not affect the distribution of the statements when we only focus on a single method. This observation shows the difficulty of inferring the privacy labels from looking at a single method. An annotator might need to refer to the method call graph or dependencies to create more precise labels. For example, an annotator may know the data is read by the third-party API by looking at the first method, but they cannot know how the data is used unless they also purview the dependent method calls. These results imply that any future research should incorporate the method dependency when creating privacy labels.

\section{Detecting Relevant Statements}
\label{sec:approach}

This section describes our approach for automatically detecting relevant statements. Essentially our approach is to fine-tune a language model to identify the first-most, second-most, and third-most relevant statements to a given privacy label in a given section of source code. We design prompts for training based on the empirical study in the previous section. The overall process is to 1) craft a prompt template, 2) create training and test datasets with prompt templates using the human study data, and 3) fine-tune a model using the prompts.

\subsection{Prompt Template}

The prompt template we follow is:

\begin{verbatim}
    CODE:\t<code>\n
    LABEL:\t<label>\n
    STATEMENT:<s>\t<statement1>
    \t<statement2>\t<statement3></s>
\end{verbatim}

The \texttt{<code>} tag is replaced with the source code for the Android method we are studying.  The \texttt{<label>} tag is replaced with the name of the ``purpose'' privacy label for that method.  The \texttt{<statement[1,2,3]>} tags are replaced with the first-most, second-most, and third-most important statements annotated by participants in the study.  The whole prompt is used for training.  During inference, we provide the following:

\begin{verbatim}
    CODE:\t<code>\n
    LABEL:\t<label>\n
    STATEMENT:<s>\t
\end{verbatim}

Then we allow the model to predict tokens until the model generates a \texttt{</s>} tag or reaches a maximum of 256 tokens.

\subsection{Training/Test Data}
\label{sec:trainingtestdata}

We divided the  2,426 annotated methods from the study in Section~\ref{sec:study} into subsets of training, validation, and test samples. We used 1,951 annotated methods as training samples and 216 samples as the validation samples. To maximize the reproducibility and reliability of results, we used the 259 methods that were annotated by two participants as the test set (see Section~\ref{sec:studymethod}). Methods in the test set are not included in validation and training sets.

\subsection{Language Models}

We frame the problem of detecting statements important to privacy labels as a fine-tuning target for a language model.  In principle, the prompt template above (or one similar to it) could be used in any language model with a sufficient context window.  We fine-tune three different language models with very different characteristics in this paper, as we seek to show how  our fine-tuning method and prompt templates apply to different types of models:

\subsubsection{Jam (350m parameter GPT-2 architecture)}
The \emph{jam}~\cite{chia2023language} language model was pretrained on a dataset of 52m Java methods and was designed to be fine-tuned for various code intelligence tasks.  The model has the advantage that its pre-training datasets are publicly available online, allowing us to check for and avoid any data contamination.  It is considered large enough to be competitive (350m parameter GPT-2 architecture), yet small enough to run on accessible hardware (a single 16GB Ampere GPU).  We followed the procedures recommended for fine-tuning \emph{jam} provided by Su~\emph{et al.}~\cite{chia2023language}.  We trained for 10 epochs, after which validation accuracy ceased improving. Model details are in table~\ref{tab:modelsetting}.

\subsubsection{GPT-3.5}

We fine-tune the 175 billion parameters \texttt{gpt-3.5-turbo-1106} model~\cite{brown2020language} using the fine-tuning API provided by OpenAI~\cite{GPTAPI}.  This model represents a very strong commercial baseline due to its good performance across multiple tasks.  It has the advantage of a high reputation for quality and very large size, but has the disadvantage of a closed dataset and missing training details.  We also have no way of knowing for sure what procedures the fine-tuning API employs.  Note that at the time of writing, we did not have access to fine-tune GPT-4.  We use the recommended settings for fine-tuning, namely letting the system decide the key training parameters automatically.

\subsubsection{CodeLlama Instruct 70B}

We also fine-tune the \texttt{codellama-instruct-70b} model provided by Meta~\cite{roziere2023code}.  This model is a balance between the previous two models.  Its size is competitive with large, commercial models such as GPT-3.5 and it is pretrained professionally with a very large (and yet closed) dataset of source code and samples of instructions from programmers.  But, we do not rely on a commercial API for fine-tuning and therefore can report training parameter details.  We use the QLora approach described by Dettmers~\emph{et al.}~\cite{dettmers2024qlora} for fine-tuning the model.  The key settings we used are in Table~\ref{tab:modelsetting}.

\begin{table}[h!]
	\centering
    \vspace{-1mm}
    \small
        \caption{Key model settings}
        \label{tab:modelsetting}
	\begin{tabular}{llll}
             &  &  Jam  & CodeLlama\\
		$g$ & gradient accumulation steps & 32 & 16 \\
		$e$ & epochs            & 10  & 4 \\
		$a$ & attention heads          & 16 & Unknown \\
		$l$ & number of layers               & 24 & Unknown\\
		$d$ & embedding dimensions                  & 1024 & Unknown \\
		$b$ & batch size                            & 4 & 1 \\
		$r$ & learning rate                         & 3e-5 & 1e-4\\   
		$d$ & dropout								& 0.2 & 0.0\\
            $o$ & optimizer								& Adam & Paged Adamw 32bit \\
            $p$ & number of parameters								& 350M & 70B \\
	\end{tabular}
    \vspace{-4mm}
\end{table}

\subsection{Software / Hardware Details}

Our hardware configuration is a workstation with an Intel i9-10900X CPU, 256GB system memory, and four NVidia A5000 GPUs (24GB video memory each).  Our software configuration includes Linux 5.15 kernel, CUDA 12.2, and PyTorch 2.0.0.  For reproducibility, see our online appendix in Section~\ref{sec:codestatement}.

\section{Evaluation}
\label{sec:eval}

In this section, we evaluate the performance of our approach for predicting the statements in the methods' source code that are most relevant to a given privacy label. We do this by analyzing the agreement between the judgment of human participants and our automated approach using each of the three language models described in the previous section.

\subsection{Research Question}
We ask two Research Questions (RQs) to evaluate the agreement between our approach and human participants:
\begin{description}
    \item [RQ3:] What is the overall agreement for relevant statements predicted by our automated approach compared to those selected by the human participants?
    \item [RQ4:] What is the agreement for the order of statements predicted by our automated approach compared to those selected by the human participants?
\end{description}

The rationale behind RQ3 is to evaluate how well our automated approach can predict the statements most likely to be selected by human participants as relevant to privacy labels. The main motivation behind our approach is to isolate code statements for developers to investigate the privacy behaviors. Therefore, we evaluate our approach by comparing our predictions against the judgment of our human participants from the web survey study described in Section~\ref{sec:study}.

The rationale behind RQ4 is to evaluate how well our approach can predict statements in the same order as selected by the participants. We asked each participant to select at most 3 statements: 1) first-most relevant, 2) second-most relevant, and 3) third-most relevant to the privacy label. We want to evaluate which of the three language models has the highest agreement with human annotators in selecting the relevant statements in the same order.

The rationale behind these two RQs is to understand how well our models performs when compared with the human experts, and how much our models assist with code statements generation given privacy labels. The model that has similar results with human experts can help the creation of the privacy labels with low cost. Note that our goal is not to compete with human experts, but to assist them in the label creation.

\subsection{Calculating Agreement}

We calculate agreement by counting the number of samples that each approach accurately predicts as the three most-relevant code statements for each sample in the test set (see Section~\ref{sec:trainingtestdata}). 
Essentially, we calculate the overlap in top-3 statement predictions between the human participants and our automated approach. Note that we do not use Cohen's Kappa or other ``agreement'' metrics on the recommendation of Delgado~\emph{et al.}~\cite{delgado2019cohen}, who argue strongly against using these metrics for classification tasks, especially in light of potentially imbalanced labels~\cite{donker1993interpretation}.

Instead, for RQ3, we compute overlap as the number of correct statements predicted as compared to human participants. For example, if participant A selected statements 2, 3, and 4 as most relevant to the privacy behavior, and our approach predicted statements 3, 4, and 5, that would be considered a two-statement overlap -- because statement 3 and 4 are both present in participant A and our approach. For RQ4, we compute overlap as the number of correct statements predicted in the same order as the human participant. In the example above, the overlap score would be zero because neither of statements 3 or 4 were predicted in the order ranked by the participant.

\subsection{Results for RQ3: Overall Agreement}

Table~\ref{tab:rq3} summarizes the results of our agreement analysis. Recall that the test set includes only methods for which we had two human annotators.  The column ``human'' refers to the overlap of these two annotators. The way to read the table is that for 5.41\% of the test set, the two human annotators agreed on all three statements irrespective of order.  For 72.20\% of samples (5.4+35.9+30.9), at least one statement overlapped in some order. We make two main observations from this data.

\begin{table*}[h]
\centering
\caption{Overlap between the order of predicted statements and the order in which human participants selected statements}
\label{tab:rq4}
\begin{tabular}{l|r|rrr|rrr|rrr|l}
Statement &      & \multicolumn{3}{c|}{Participant A}    & \multicolumn{3}{c|}{Participant B} & \multicolumn{3}{c}{Average}\\
Overlap   & human & jam     & gpt-3.5       & codellama & jam & gpt-3.5       & codellama & jam & gpt-3.5 & codellama &\\ \cline{1-11}
three     & 4.63\%                    & 5.02\%  & 1.93\%        & 3.47\%    & 5.41\%                  & 3.86\%         & 3.09\%                        & \textbf{5.21\%}                 & 2.90\%                      & 3.28\% & higher is better \\
two       & 6.18\%                    & 8.11\%  & 4.25\%        & 9.27\%    & 6.56\%                  & 7.34\%                      & 5.79\%                        & 7.34\%                  & 5.79\%                      & 7.53\% &                       \\
one       & 16.22\%                   & 16.60\% & 15.44\%       & 14.67\%   & 20.08\%                 & 11.97\%                     & 15.83\%                       & 18.34\%                 & 13.71\%                     & 15.25\% &                      \\
zero      & 72.97\%                   & 70.27\% & 78.38\%       & 72.59\%   & 67.95\%                 & 76.83\%                     & 75.29\%                       & \textbf{69.11\%}                 & 77.61\%                     & 73.94\%  & lower is better                     
\end{tabular}
\vspace{-3mm}
\end{table*}

\begin{table*}[t!]
\centering
\caption{Overlap between predicted statements and statements selected by human participants.}
\label{tab:rq3}
\begin{tabular}{l|r|rrr|rrr|rrr|l}
Statement &      & \multicolumn{3}{c|}{Participant A}    & \multicolumn{3}{c|}{Participant B} & \multicolumn{3}{c}{Average}\\
Overlap   & human & jam     & gpt-3.5       & codellama & jam & gpt-3.5       & codellama & jam & gpt-3.5 & codellama & \\ \cline{1-11}
three     & 5.41\%                    & 6.18\%  & 6.18\%        & 10.81\%   & 5.79\%                  & 6.56\%                            & 5.41\%                        & 5.98\%                  & 6.37\%                      & \textbf{8.11\%} & higher is better               \\
two       & 35.91\%                   & 32.05\% & 31.27\%       & 28.96\%   & 36.29\%                 & 26.64\%                           & 36.29\%                       & 34.17\%                 & 28.96\%                     & 32.63\%        &               \\
one       & 30.89\%                   & 34.75\% & 33.98\%       & 29.73\%   & 37.07\%                 & 39.00\%                           & 31.27\%                       & 35.91\%                 & 36.49\%                    & 30.50\%     &                  \\
zero      & 27.80\%                   & 27.03\% & 28.57\%       & 30.50\%   & 20.85\%                 & 27.80\%                           & 27.03\%                       & \textbf{23.94\%}                 & 28.19\%                     & 28.76\%  & lower is better                   
\end{tabular}
\vspace{-0.2cm}
\end{table*}

First, we found that there is a high degree of subjectivity in human judgment about which statements are important for a privacy label. For 27.8\% of samples, the annotators did not agree on any of the three statements. This may be because we could not verify the background of participants in the privacy domain and they may have varying degrees of experience although some works indicate that even privacy experts disagree \cite{li2022understanding, balebakoyour2014}. This high degree of disagreement shows the difficulty of the problems based on a single method as discussed in Section~\ref{sec:implication}

Second, we observed that although~\emph{jam} is the smallest language model of the three, it is better than the two bigger models at predicting at least some overlap. We observe that \emph{jam} has the smallest sample size with a zero overlap when compared with either Participant A (27.03\%) or Participant B (20.85\%). Surprisingly, on an average between the two participants, \emph{jam} approach predicts at least 1 label correctly irrespective of order for 76.06\% of samples, compared to 71.81\% for \emph{gpt-3.5} and 71.24\% for \emph{codellama}. The \emph{codellama} approach seems to excel for higher $overlap=3$ with an agreement of 10.81\%.  We posit that the larger sized \emph{gpt-3.5} and \emph{codellama} may be performing poorly on this domain-specific task because of the difference in fine-tuning strategies employed for our task (see Section~\ref{sec:approach} and~\ref{sec:threats}). Another factor may be prompt design, as we have not extensively performed prompt engineering for our task. Prompt variants and its impact on the large LLMs may be tackled in the future.

Overall, we found that \emph{jam}, the smaller of the language models performs better than the larger language models for our task. We also observe that all language models show higher agreement with participant B than with participant A, and the agreement between the language model and a participant is higher compared to the agreement between participants themselves. One reason for this could be varying levels of experience with privacy labels among the participants.

\subsection{Results for RQ4: Agreement in Order}

Table~\ref{tab:rq4} summarizes the results of our agreement analysis for RQ4 when considering the order of the statements. First, we found that the agreement between human annotators is much lower than Table~\ref{tab:rq3}. For 72.97\% of the samples, the two human annotators did not agree on even one statement in the same order of relevancy to privacy behavior. The agreement between human annotators is lower than both \emph{jam} and \emph{codellama}. This observation is similar to RQ3 and further suggests a high degree of variance in human judgment. 

Next, we found that similar to RQ3, our approach with \emph{jam} achieves a higher agreement with human annotators compared to \emph{codellama} and \emph{gpt-3.5}, and even is higher than the agreement between human annotators. Surprisingly, contrary to the findings in RQ3, \emph{jam} also outperforms \emph{codellama} at $overlap=3$, as well. These observations further support our notion that for domain-specific tasks like this, smaller domain-specific language models may outperform general-purpose language models like \emph{gpt-3.5} (in instances without extensive prompt engineering). Another advantage of domain-specific open source models like \emph{jam} is the ability to check if the model has already seen the data during training~\cite{chia2023language}.

Overall, based on the results of RQ3 and RQ4, our automated approach shows a higher agreement with individual human participants compared to the agreement between these participants themselves, both with and without considering order. We found that smaller language models like \emph{jam} outperform larger commercial models for this specific task of identifying code statements most relevant to privacy behaviors. In light of this finding, we recommend not ruling out smaller domain-specific language models over larger sized models for focused tasks, such as ours, which may be a representative of several other tasks relating to automatic code comprehension and classification of code statements.

\begin{figure}[h!]
    \centering
    \includegraphics[scale=0.25]{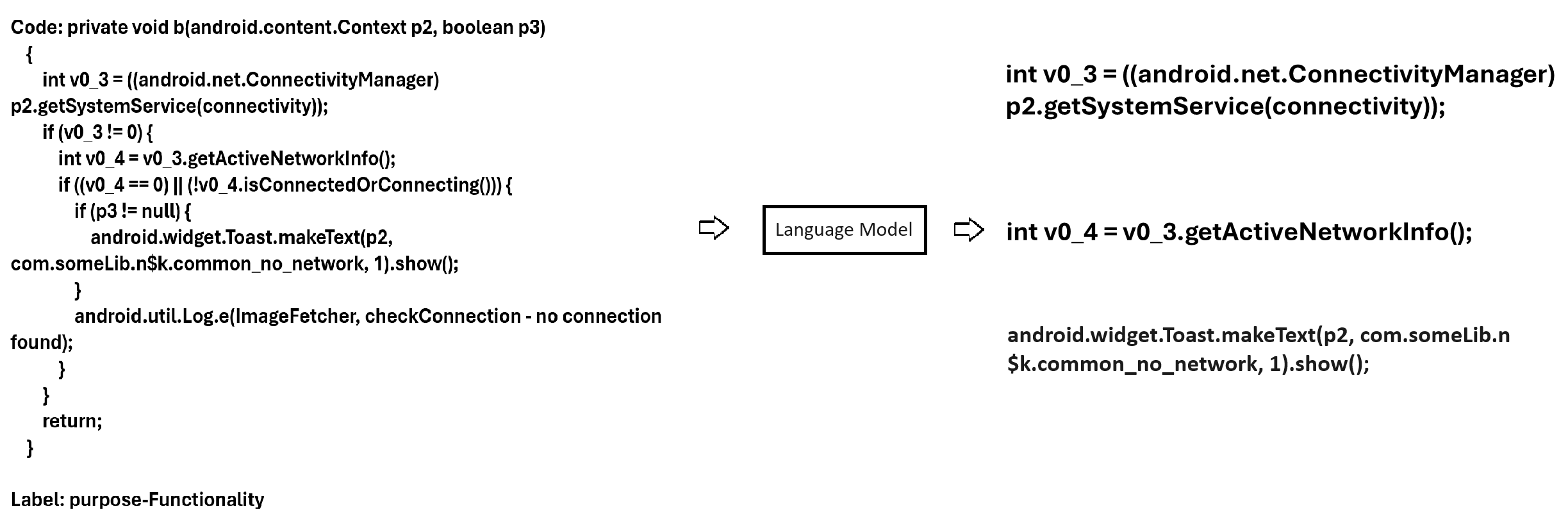}
    \vspace{-1mm}
    \caption{An example of our study findings. The left sub-figure shows the source code of an Android method from our dataset, and the right sub-figure highlights the three statements predicted by our approach based on the privacy label.}
    \label{fig:output_example}
    \vspace{-5mm}
\end{figure}

\begin{figure}[b!]
\vspace{-0.3cm}
\centering
\begin{tabular}{c}
\includegraphics[width=0.5\linewidth]{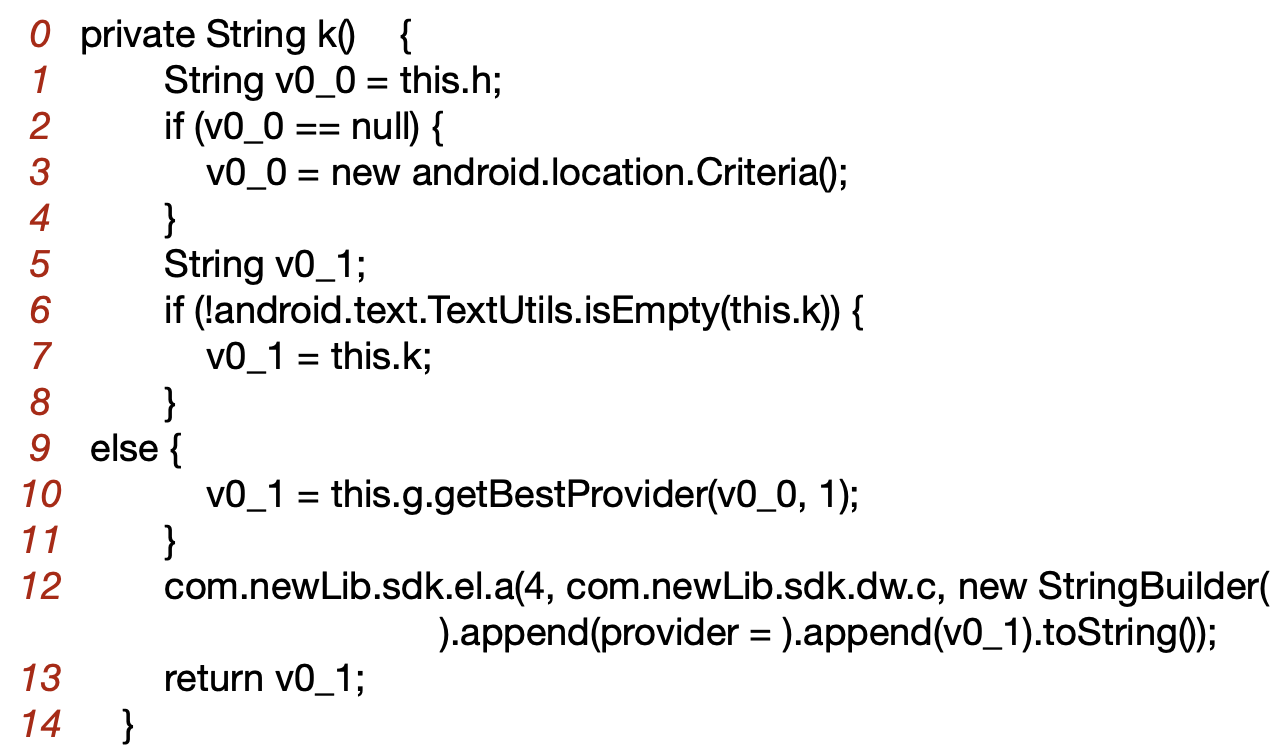}\\
\end{tabular}
\begin{tabular}{l|lllll}
Order  & Part-A & Part-B & jam & gpt-3.5 & codellama \\ \hline
first-most  & 3      & 1      & 6            & 0                & 12                 \\
second-most & 7      & 6      & 10           & 12               & 5                  \\
third-most  & 12     & 12     & 12           & 13               & 6   \\ 

\multicolumn{6}{c}{~} \\
Overlap A & & & & & \\ \hline
any order  & x    & 1     & 1           & 1               & 1   \\
in order & x    & 1     & 1           & 0               & 0   \\
\multicolumn{6}{c}{~} \\
Overlap B & & & & & \\ \hline
any order  & 1     & x     & 2           & 1               & 2   \\
in order & 1     &  x    & 1           & 0               & 0   \\
\multicolumn{6}{c}{~} \\
\end{tabular}
\caption{An example of our study findings. Sub-figure on the left shows the source code of an Android method from our dataset, with code statements, indicated by line numbers in red. Sub-figure on the right presents sentences in order of relevancy selected by two participants A and B, as well as the ones predicted by our approach using three language models. We anonymized the name of the project as `newLib'.}
\label{fig:example}
\end{figure}

\section{Example \& Discussion}
In this section, we discuss a few examples showcasing the statements selected by our participants and the outputs predicted by our approach. We aim to illustrate motivation and the efficacy of our proposed approach in these examples.

\subsection{Output Example from our Approach}
Figure~\ref{fig:output_example} shows the output example from our approach. Given the source code and the privacy label, our approach predicts three statements to be relevant. This fine-grained output helps the stakeholder quickly understand the privacy-relevant aspects of the source code. When there is a source code update, the stakeholder can easily monitor changes to the privacy-relevant statements and update the related privacy labels. On the other hand, a coarse-grained analysis that is not linked with the source code would require the stakeholder to scrutinize the entire source code for changes to privacy-relevant aspects.

\subsection{Example on Efficacy of our Approach}
Figure~\ref{fig:example} shows an example of (a) an Android method from our dataset, and (b) a table comparing statements selected by two participants and those predicted by our approach with three LLMs. Additionally, the table shows max overlap between the selected and predicted statements. Note, we report the line numbers correlating to the statement to better present results, but the participants saw raw code to select statements on the web-survey and the approach generates code statements as a sequence of word tokens.

Columns Part-A and Part-B report the statements selected by two participants in our human study. The two participants show disagreement for the~\emph{first-most} and~\emph{second-most} relevant statements. However, they both agree on the~\emph{third-most} relevant statement. In this instance, the max overlap between them is one, both with and without considering order. This observation supports our findings in Section~\ref{sec:eval} that the judgement of participants is highly variable.

For the automated approaches, we found that the~\emph{jam} model shows a higher agreement with Part-B on statements 6 and 12. However,~\emph{jam} predicts statement 6 to be~\emph{first-most} relevant statement. The max overlap between~\emph{jam} and Part-B is two but if the order is considered the max overlap is one.~\emph{codellama} also has a max overlap of two, but if the order is considered the max overlap is zero. As observed in Section~\ref{sec:eval},~\emph{gpt-3.5} achieves a low overlap score of one (any order) and zero (in order), when compared to either of the participants. While this is just one sample, we chose this example to help illustrate our earlier findings.

\vspace{1mm}
\section{Threats to Validity}
\label{sec:threats}
Like any experiment, this study carries threats to its validity.  Key threats include the interface design, the source code we selected, the selection of participants in our human study, and the language models we use for the automated approach.

Any flaws in the interface design could lead people to provide incorrect responses.  We attempted to avoid major biases by not implying that we had an investment in the outcome of the experiment (demand characteristic bias~\cite{dell2012yours}), conducting thorough pilot studies with people whose results we did not include in this experiment, and the addition of the option for no relevant line (upon feedback in pilot studies).  

The source code we sampled for the study could have influenced the findings, because different methods may result in different selections and therefore different conclusions.  Our mitigation of this threat was to use a representative sample with 2,426 methods, randomly chosen from a published and vetted dataset of privacy labels~\cite{jain2022pact}.  Nonetheless, we caution that our conclusions may only be meaningful for Android methods of approximately the same size and composition.  

The participants in our study also influence our results, mainly because we conduct this study on a web platform. We could not confirm the participant's background or experience with Android app development or knowledge in privacy.  We took measures to mitigate this by screening participants based on their replies, and discarding data from participants with nonsensical rationales or statement selections. Participant opinions are also subjective, so we caution that our results may not represent the opinions of a large set of developers.

The chosen language models influence the results. Because two of the models are commercial in nature, we cannot control some of their parameters -- such as the API call process used to finetune~\emph{GPT-3.5}~\cite{GPTAPI} and the low-rank adapter LoRA used to finetune~\emph{codellama}~\cite{dettmers2024qlora}. Therefore, we re-iterate that our results in Section~\ref{sec:eval} (that compares LLMs) only applies to our specific problem and setup, and are not an exhaustive or objective evaluation of the capabilities of these language models. Moreover, the chosen prompt string has a strong influence on the performance of these closed language models. Performing extensive prompt engineering specific to each model may produce different results.
\section{Conclusion}
\label{sec:conclusion}
In conclusion, we make the following key contributions and findings in this paper.
We conducted an empirical web-survey style study on the Prolific platform, where 18 programmers identified the three mode relevant code statements in an Android method that are most relevant to a given privacy label. 
Our evaluation showed that participants identified expression and declaration type code statements (that make function calls, and augment or save the data to local variables) as most relevant to privacy behaviors. The participants also identified `if' conditional statements to be of secondary relevance to the privacy behaviors. 
We found that the `purpose' of the privacy behavior, as described by the privacy label, has very little effect on the type of code statements identified by the participants as relevant. 
We presented a neural language model based approach to automatically identify the privacy relevant code statements given an Android method and a privacy label. We use three language models to test our approach, namely~\emph{jam}~\cite{chia2023language},~\emph{GPT-3.5}~\cite{GPTAPI}, and~\emph{codellama}~\cite{roziere2023code}.
We evaluated our automated approach by comparing the predicted statements against reference selected by the participants for overlap and found that for our task, the smallest of the three language models~\emph{jam} outperforms bigger language models. We also found that human participants show a high degree of disagreement, 
and our automated approach shows a higher agreement with individual human participants compared to the agreement between the participants themselves.

\section*{Acknowledgments}
This work is supported in part by NSF CCF-2100035, CCF-2211428, CCF-2238047. Any opinions, findings, and conclusions expressed herein are the authors and do not necessarily reflect those of the sponsors.

\section*{Conflict of Interest}
The authors declare no potential conflict of interests.

\section*{Data/Code Availability} 
\label{sec:codestatement}
We release our data and code for experiments in our Github repository, 
\url{https://github.com/apcl-research/prigen_statement}





\bmsection*{Acknowledgments}
This work is supported in part by NSF CCF-2100035 and CCF-2211428. Any opinions, findings, and conclusions expressed herein are the authors and do not necessarily reflect those of the sponsors

\bmsection*{Conflict of interest}

The authors declare no potential conflict of interests.

\bibliography{wileyNJD-AMA}



\end{document}